\documentclass[pre,twocolumn,showpacs,floatfix,nofootinbib,groupaddress,showkeys]{revtex4}
\usepackage{graphicx}
\usepackage[dvips]{color}
\usepackage{epsfig}
\usepackage{amssymb}
\usepackage{amsmath}
\usepackage{bm}

\begin{document}

\bibliographystyle{apsrev}
\title{Defect-mediated emulsification in two dimensions}
%\author{K.S.~Korolev and David R. Nelson}
\author{K.S.~Korolev}
\email{papers.korolev@gmail.com}
\affiliation{Department of Physics, Harvard University, Cambridge, Massachusetts 02138, USA}
\author{David R. Nelson}
\affiliation{Department of Physics, Harvard University, Cambridge, Massachusetts 02138, USA}
\begin{abstract}

%\normalsize
We consider two-dimensional dispersions of droplets of isotropic phase in a liquid with an \textit{XY}-like order parameter, tilt, nematic, and hexatic symmetries being included. Strong anchoring boundary conditions are assumed. Textures for a single droplet and a pair of droplets are calculated and a universal droplet-droplet pair potential is obtained. The interaction of dispersed droplets via the ordered phase is attractive at large distances and repulsive at short distances, which results in a well defined preferred separation for two droplets and topological stabilization of the emulsion. This interaction also drives self-assembly into chains. Preferred separations and energy barriers to coalescence are calculated, and the effects of thermal fluctuations and film thickness are discussed.

\end{abstract}

\date{\today}
%\smallskip

\maketitle

%\large
\section{\label{SIntro}Introduction}

Emulsions are an important nonequilibrium state of bulk matter, widely used in the industry and of intrinsic physical interest. Although two-dimensional emulsions are less common, they may mediate certain membrane bounded process in living cells~\cite{Simons,Veatch,Baumgart1,Baumgart2}, control the kinetics of phase transitions in thin films, and shed light on the corresponding phenomena in three dimensions.

One of the main questions in studies of emulsions from both a theoretical and experimental point of view is stabilization. Since emulsions are out of equilibrium, they tend to phase separate. Therefore for an emulsion to be useful it has to be stabilized by kinetic barriers to coalescence. Typical strategies include surfactant or grafted polymer stabilization and could involve Coulombic or steric repulsion of dispersed droplets~\cite{Larson,Israelachvili}. Recently there has been substantial interest in topological stabilization, which may arise when droplets of disordered liquid are introduced in an anisotropic host fluid~\cite{Poulin,Lubensky,Pettey2,Bohley,Cluzeau1,Voltz,Cluzeau2,Dolganov,Fukuda,Yamamoto,Tasinkevych,Patricio,Fukuda0,Fukuda1,Stark0,Yamamoto1,Loudet,Nazarenko,Anderson}.

Anisotropic liquids exist both in two and three dimensions. Tilt and bond orientational order are quite common in liquid monolayers~\cite{Knobler}, and there is some evidence that hydrated lyotropic bilayers of DMPC~(dimyristol-phophatidylcholine, saturated fatty acids chains) have bond orientational order as well~\cite{Smith,Chiang}. Quite recently, a new hexatic phase, with six-fold orientational order, has been discovered in freestanding lyotropic thin films as few as four bilayers thick~\cite{Chev}.

Langmuir-Blodgett monolayers at air-water interfaces have long lived domains due to long-range electric dipole interactions between lipids at the asymmetric air-water interface~\cite{Andelman}. However, these dipole interactions are much reduced for \textit{bilayer} systems in water, and are unlikely to account for the arrested phase separation observed in, e.g., Refs.~\cite{Baumgart1,Langer}. Komura~\textit{et. al}. have constructed a mean field theory which illuminates a number of aspects of lateral phase separation in mixtures of lipid and cholesterol in bilayers~\cite{Komura}. The ``lipid ordered'' and gel phases are described by a simple Ising-type order parameter without, however, explicit regard for the orientational order that may be responsible for the slow coalescence.

Two-dimensional topological emulsions have also been observed experimentally in thin films of smectic-\textit{C/C*} liquid crystal~\cite{Cluzeau1,Voltz,Cluzeau2,Dolganov}. In such systems the order parameter is the projection of the tilted director on the plane of the film and inclusions are typically disklike ``bubbles'' of less ordered phase (nematic or cholesteric) or domains with a different number of smectic layers. Under appropriate conditions, these inclusions are accompanied either by a topological defect in the bulk of the host fluid~\cite{Cluzeau1,Cluzeau2} or by several surface defects~\cite{Voltz,Dolganov}. The distortions in the texture lead to long-range attraction and short-range repulsion between bubbles, and they drive self-assembly into chainlike and latticelike structures~\cite{Cluzeau1,Voltz,Cluzeau2,Dolganov}. These phenomena have been studied theoretically before~\cite{Pettey2,Bohley,Fukuda,Tasinkevych,Patricio}. Analytical solutions for the texture around a \textit{single} droplet have been calculated in Refs.~\cite{Pettey2,Fukuda,Tasinkevych} and for the far-field interactions in Refs.~\cite{Pettey2,Tasinkevych}. Textures around several inclusions with surface defects have been obtained numerically by Bohley and Stannarius~\cite{Bohley}. For bulk defects, short range repulsion and stability have been discussed in Refs.~\cite{Tasinkevych,Patricio}. 

In this paper we discuss how orientational order and point defects give rise to the rich behavior seen in the experiments. We first review the basic properties of topological emulsions in Sec.~\ref{STE}. In the next three sections we verify, extend and correct the results from Refs.~\cite{Pettey2,Tasinkevych,Patricio}. In particular, we consider droplets of \textit{different} sizes embedded in a two-dimensional liquid crystal with an \textit{arbitrary} orientational order parameter. In Sec.~\ref{SSD}, we calculate textures induced by bubbles of disordered phase in a liquid with a generalized \textit{XY}-like order parameter (tilt, nematic, and hexatic order are special cases; the relatively unexplored case of~$4$-fold ``tetratic'' order is included as well.). In Sec.~\ref{SFFI}, we obtain an analytical expression for far-field interactions between bubbles mediated by the aforementioned order parameters. We then discuss numerical computation of an effective droplet-droplet pair potential in Sec.~\ref{SNCPI} and show that this interaction leads to bubble stabilization in the presence of the ``topological quasi-long range order''~\cite{Kosterlitz} characteristic of two-dimensional~(2D) systems with an~\textit{XY}~order parameter. In Sec.~\ref{SChaining} we use our numerical approach to address chaining. Finally, in Sec.~\ref{STF} we discuss how our results are affected by thermal fluctuations and film thickness.

\section{\label{STE}Topological Emulsions}

Phenomena of interest in this paper can arise in a variety of physical systems~\cite{Cluzeau1,Voltz,Cluzeau2,Dolganov,Baumgart1,Baumgart2,Smith,Chiang,Chev}. In this section we outline a model that captures many essential features and has a virtue of simplicity.

 A natural way to create a two-dimensional emulsion is to rapidly quench the system below a mixing-demixing transition. The defect-mediated emulsification of interest to us here arises when the continuous phase has order parameter with \textit{XY} symmetry. The system with both mixing-demixing and \textit{XY} ordering transitions can be described with the following coarse-grained lattice gas Hamiltonian~\cite{Berker},

\begin{equation}
\begin{split}
\label{CoarseGrainedHamiltonian}
-\frac{H}{k_{B}T}=&\frac{J}{2}\sum_{\langle i,j\rangle}t_{i}t_{j}(\psi^{*}_{i}\psi_{j}+\psi_{j}^{*}\psi_{i})+K\sum_{\langle i,j\rangle}t_{i}t_{j}-\mu\sum_{j}t_{j},
\end{split}
\end{equation}

\noindent where~$K>0$~and~$J>0$~describe interactions, $\mu$~is a chemical potential,~$i$~and~$j$~index lattice sites, ${\langle i,j\rangle}$~means a sum over nearest neighbors,~$t_{i}=0$ denotes a patch of disordered phase and~$t_{i}=1$ denotes an ordered phase region. The continuous \textit{XY}-like symmetry is embodied in the complex~$p$-fold symmetric order parameter~$\psi_{j}=e^{ip\theta_{j}}$.

\begin{figure}
\includegraphics[width=\columnwidth]{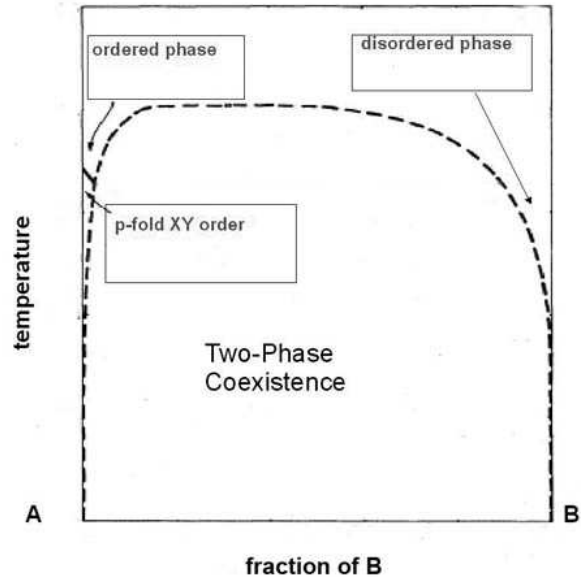}
\caption{Phase diagram for a mixture of ordered and disordered phases.}
\label{PhaseDiagram}
\end{figure}

At~low~$K$~(high temperature) entropy dominates and the system is mixed, while at high~$K$~(low temperature) it phase separates. In addition to this Ising-type term the Hamiltonian includes \textit{XY}-like interactions between sites that are occupied by the ordered phase. Here~$p=1,2$,~and~$6$ corresponds to tilt, nematic, and hexatic order parameters respectively; tetratic order is described by~$p=4$. Fig.~\ref{PhaseDiagram} shows a phase diagram obtained via real space renormalization methods for this Hamiltonian when~$K=J$~\cite{Berker}. Thermal fluctuations drive the ordering transition for the continuous symmetry degrees of freedom~$\left\{\psi_{j}\right\}$ well below the Ising-type critical point, and also lead to the broad flat top seen in experiments~\cite{Veatch,Baumgart1} and typical of the small Ising-type coexistence curve exponent~$\beta=1/8$. Because thermal fluctuations are strong in two dimensions, the thin sliver of $p$-fold \textit{XY} ordered regime on the left side only exhibits ``quasi-long-range order,'' with algebraic decay of correlations in~$\psi_{i}$~\cite{Kosterlitz}.

Well below both phase transitions the free energy of the ordered phase can be described by a gradient expansion in the angle field~$\theta_{j}$. For~$p=1,2$ we make the one elastic constant approximation,

\begin{equation}
\label{GradientExpansionEnergy}
F=\frac{g}{2}\int d^{2}x(\protect\bm{\nabla} \theta)^{2}.
\end{equation}

\noindent This representation of the low energy physics is exact for~$p>2$. Here~$g$ is the stiffness constant,~and the angle~$\theta$ is the orientational order parameter.
Functional minimization of~Eq.~(\ref{GradientExpansionEnergy}) leads to a simple Euler-Lagrange equation for~$\theta$,

\begin{equation}
\label{LaplaceEquation}
\protect\bm{\nabla}^{2}\theta=0.
\end{equation}

One important requirement for topological stabilization of emulsions is the anchoring boundary energy, which depends on the angle between the director and local normal to the boundary of a bubble. Here we assume the limit of strong boundary conditions, when this angle is chosen to minimize the boundary energy~\cite{Poulin,Lubensky,Pettey2,Bohley,Cluzeau2}. Tangential and homotropic boundary conditions are two examples when the angle is~$\pi/2$~and~$0$ respectively. Strong anchoring implies that the order parameter uniformly changes by~$2\pi$ along a bubble perimeter. For simplicity we adopt homotropic boundary conditions but the results are valid for all other anchoring angles, because the textures only differ by a rotation of the order parameter by an anchoring angle at every point in space; this rotation does not affect the energy of the system in the one elastic constant approximation of~Eq.~(\ref{GradientExpansionEnergy}).

Boundary conditions at the outer boundary of the emulsion are also very important. We assume free boundary conditions, which simplify the calculations and are appropriate when there is no anchoring at the outer boundary. In the experiments with smectic films, the director at infinity is not free but is usually forced to point in some particular direction~\cite{Cluzeau1,Cluzeau2,Voltz,Dolganov,Pettey2}. Although it would be straightforward to repeat our calculations with these boundary conditions, the difference is often unimportant because, when topological defects are included, both boundary conditions lead to uniform director field at infinity and, therefore, to the same solution up to a rotation relative to the outer boundary. However, uniform boundary conditions at infinity must be used if one studies the dependence of interaction energy on the angle~$\vartheta$ between the line connecting the droplets and the orientation of the director at the outer boundary. The free boundary conditions correspond to the angle~$\vartheta$ that minimizes the energy of the system.

\section{\label{SSD}Single Droplet}

First we consider a single circular droplet of isotropic phase with radius~$a$ embedded in a two-dimensional liquid crystal with p-fold symmetric \textit{XY} order. We assume a strong line tension preserves the circular shape. In the absence of topological defects, the only solution satisfying~Eq.~(\ref{LaplaceEquation}) and the boundary condition is~$\theta=\phi_{0}$, where~$\phi_{0}$ is the polar angle relative to the center of the droplet. Thus, from the outside, the droplet looks like a point defect with charge~$+1$. From Eq.~(\ref{GradientExpansionEnergy}) we find that the free energy of this texture is proportional to~$\pi g\ln(L/a)$, where~$L$~is the system size. However, the free energy can be lowered significantly below this logarithmic divergence if one allows for an appropriate number of point defects, so that the net charge of the droplet and the defects is zero.

The order parameter we are considering can be represented by an arrow with directions differing~by~$2\pi/p$ identified. This order parameter allows for point defects with topological charge~$n/p$, where~$n$~is an integer, which means that the order parameter rotates by~$2\pi n/p$ on any contour surrounding the defect. To make the free energy of the system finite in the presence of a droplet, we need the charges of the defects to~sum~up~to~$-1$. Upon noting that the core energy scales as~$n^{2}$ and that many defects with smaller charges screen better than few defects with larger charges, we conclude that~$n=1$ and study~$p$~defects with charge~$-1/p$. From the symmetry considerations and because defects repel each other it is clear that they must form a regular $p$-gon around the droplet, at least for~$p$ of physical interest,~$p \le 6$. This behavior was also found numerically by Fukuda and Yokoyama~\cite{Fukuda1} for~$p=2$ considered there.

To completely specify equilibrium locations of the defects we have only to determine the distance~$R$ between the center of the droplet and the defects. On dimensional grounds,~$R=\delta a$, where~$\delta$~is dimensionless. To calculate~$\delta$ we have to find a solution of~Eq.~(\ref{LaplaceEquation}) that satisfies the boundary condition and has appropriate singularities for the pattern of defects specified above. This is most easily accomplished by the method of images. The solution is

\begin{equation}
\theta=2\phi_{0}-\frac{1}{p}\sum_{i=1}^{p}(\phi_{i}+\phi'_{i})+\rm{const},
\label{SingleDropletSolution}
\end{equation}

\noindent where the~$\{\phi_{i}\}$~are polar angles around the defects and the~$\{\phi'_{i}\}$ are polar angles around the images obtained by inversion in the circle. Technical details of the derivation are discussed in Appendix~\ref{ASD}. The inverse problem of textures \textit{inside} a circular droplet of ordered phase embedded in a disordered isotropic phase is discussed in Appendix~\ref{ASDI}.

From Eqs.~(\ref{SingleDropletSolution})~and~(\ref{GradientExpansionEnergy}) we obtain the free energy as a function of~$R$, namely,

\begin{equation}
    F=\frac{\pi g}{p}\ln\left(\frac{R^{3p+1}}{p a^{p}(R^{2p}-a^{2p})c}\right)+pE_{c},
\label{SingleDropletEnergy}
\end{equation}

\noindent where~$c$ is the core radius and~$E_{c}$~is the core energy. Upon minimization, Eq.~(\ref{SingleDropletEnergy}) gives~(with~$\delta=R/a$)

\begin{equation}
    \delta=\left(\frac{3p+1}{p+1}\right)^{\frac{1}{2p}},
\label{Delta}
\end{equation}

\noindent and a free energy that no longer diverges with the system size~$L$,

\begin{equation}
	F=\frac{\pi g}{p} \ln\left(\frac{\delta^{3p+1}}{p(\delta^{2p}-1)}\frac{a}{c}\right)+pE_{c}.
\end{equation}

Our results agree with the calculation in Ref.~\cite{Pettey2} and experiments by Cluzeau \textit{et al.}~\cite{Cluzeau1}, which apply to the case~$p=1$ and with the calculation in Ref.~\cite{Tasinkevych}, which apply to the case~$p=2$; they are also consistent with the numerical work discussed in Sec.~\ref{SNCPI}. In another experimental system with~$p=1$, a different texture has been observed~\cite{Voltz,Dolganov}. Instead of one topological defect in the bulk, a droplet has two and sometimes four \textit{surface} defects of strength~$1/2$~and~$1/4$ respectively. In the case of the one elastic constant approximation and free boundary conditions at infinity considered here, such defects are located equidistantly along the circle because they repel each other. The texture could be obtained by the method discussed in Appendix~\ref{ASD}; in this case the defects and images are located at the same points. For two half-defects the elastic energy is decreased by~$3\ln(2)\pi g$ compared to the texture with one defect outside the bubble, but the anchoring energy is increased because the boundary conditions must be violated around two singular points. The energy difference between the alternative textures does not depend on the size of the droplet and the winning configuration is determined by the competition between elastic and anchoring energies. We assume here anchoring energies large enough to exclude surface defects on the bubble.

In summary, we have obtained the exact texture around a single droplet with strong anchoring boundary conditions on the surface of the bubble and free boundary conditions at infinity. One important conclusion is that, in addition to line tension and bulk contributions to the droplet free energy, there is a new size-dependent contribution to the droplet free energy~$(\pi g/p)\ln(a/c)$. This additional energy will alter the usual estimates for nucleation and growth of droplets of isotropic phase in terms of bulk and surface energies.

It is also important to emphasize that the penalty for violating the strong anchoring boundary condition on the droplet itself, by superimposing, say, a uniform texture around the bubble, grows linearly with~$a$. However, the creation of several topological defects only bears the fixed cost of the core energies. Thus, small droplets may not be accompanied by the defects, while large ones are.

While we have shown that a single droplet acquires companion defects, demonstrating topological stabilization of emulsions requires more. We now consider the interaction of two such domains as a function of the center-to-center distance~$r$ between them, and show that it leads to large barriers to coalescence.

\section{\label{SFFI}Far-field interactions}

Similar to our analysis of a single droplet, we expect to have~$2p$ defects with charges~$-1/p$ to neutralize the topological charge of \textit{two} droplets with strong anchoring boundary conditions. Unlike the single droplet case, we cannot predict the location of the defects based only on the symmetry; we need other methods to find equilibrium positions of the singularities for each droplet separation~$r$ to compute an effective droplet-droplet pair potential~$V(r)$.

\begin{figure}
\includegraphics[width=\columnwidth]{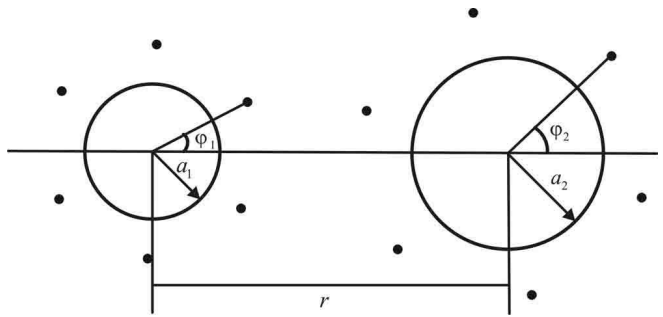}
\caption{Geometry of the problem in the far field limit,~$r\gg a_{1},a_{2}$. Droplets of radius~$a_{1}$~and~$a_{2}$ are surrounded by regular $p$-gons of defects, and the pair potential depends only on their orientations and the separation~$r$.}
\label{TwoDropletsSetup}
\end{figure}

In the far-field limit, when the distance between the droplets is much greater than their sizes, one can still assume that the defects form a regular $p$-gon around each of the droplets; the sizes of these $p$-gons remain unaffected, to leading order in~$a/r$. What one cannot neglect is the relative orientation of the $p$-gons. Due to the rotational symmetry of an isolated droplet, its energy does not depend on relative orientation of the $p$-gon and the droplet. For two droplets the symmetry is reduced, and the interaction should set the preferred orientations of the $p$-gons. The geometry of the problem is summarized in Fig.~\ref{TwoDropletsSetup}.

Thus, we have to solve~Eq.~(\ref{LaplaceEquation}) subject to the homotopic boundary conditions used for a single droplet and specified locations of the point defects. The calculation is not straightforward because~$\theta$ is defined only up to~$2\pi/p$ rotations. This leads one either to consider multivalued functions or to introduce many branch cuts, both of which complicate the problem. An alternative approach is to introduce auxiliary single valued fields~$\psi$~and~$\chi$ such that,

\begin{equation}
\label{PsiChiDecomposition}
\protect\bm{\nabla}_{i}\theta=\protect\bm{\nabla}_{i}\psi-\bm{\epsilon}_{ij}\protect\bm{\nabla}_{j}\chi,
\end{equation}

\noindent where~$\bm{\epsilon}_{ij}$ is the antisymmetric tensor in two dimensions. It is possible to decouple the auxiliary fields by demanding that~$\psi$ is constant on the surface of the droplets (the constants do not have to be the same on different droplets), which implies that~$\protect\bm{\nabla}\psi$ is normal to the boundaries. The energy is then given by

\begin{equation}
\label{NewEnergy}
F=\frac{g}{2}\int(\protect\bm{\nabla}\psi)^{2}d^{2}x+\frac{g}{2}\int(\protect\bm{\nabla}\chi)^{2}d^{2}x,
\end{equation}

where the cross term vanishes since

\begin{equation}
\label{NoCrossTerm2}
\begin{split}
-\int(\bm{\epsilon}_{ij}\protect\bm{\nabla}_{i}\psi\protect\bm{\nabla}_{j}\chi)d^{2}x=&\int(\chi\bm{\epsilon}_{ij}\protect\bm{\nabla}_{i}\protect\bm{\nabla}_{j}\psi)d^{2}x+\\&\int(\chi\bm{\epsilon}_{ij}n_{j}\protect\bm{\nabla}_{i}\psi)dl.
\end{split}
\end{equation}

The first term on the right hand side is zero due to antisymmetry of~$\bm{\epsilon}_{ij}$. The second term vanishes for a similar reason, because we choose~$\protect\bm{\nabla}\psi$ to be parallel to the outward unit normal to the boundary~$\mathbf{n}$. We also assume that~$\psi$~and~$\chi$ decay fast enough that there is no contribution at infinity.

It follows from Eq.~(\ref{NewEnergy}) that both~$\psi$~and~$\chi$ satisfy Laplace's equation. One can also show that topological defects in the~$\theta-$field correspond to quantized singularities in~$\chi-$field and that for any texture there always exist~$\psi$~and~$\chi$ which satisfy~Eq.~(\ref{PsiChiDecomposition}). The approach of dividing the gradient of the order parameter in two fields is similar to the electromagnetic analogy exploited in Ref.~\cite{Pettey2}.

To find the interaction of droplets we solve the following set of equations:

\begin{equation}
\label{EquationPsiChi1}
\protect\bm{\nabla}^{2}\psi=0,
\end{equation}

\begin{equation}
\label{EquationPsiChi2}
\protect\bm{\nabla}^{2}\chi=\frac{2\pi}{p}\sum_{i=1}^{2p}\delta^{(2)}(\mathbf{x}-\mathbf{d_{i}}),
\end{equation}

\noindent where~$\mathbf{x}$ labels points in space and~$\mathbf{d_{i}}$ refers to the positions of the defects. We choose boundary condition for~$\chi$ on the surface of the first droplet to be

\begin{equation}
\label{BoundaryConditionsChi}
\protect\bm{\nabla}\chi\centerdot\mathbf{\hat{n}}=1/a_{1}.
\end{equation}

\noindent This condition guarantees that the order parameter uniformly changes along the boundary by~$2\pi$. However, the director generally forms a nonzero angle~$\alpha_{1}$ with the normal. This discrepancy can be easily fixed by imposing a boundary condition on~Eq.~(\ref{EquationPsiChi1}) for~$\psi$ that effectively rotates the order parameter to match the normal as follows:

\begin{equation}
\label{BoundaryConditionsPsi}
\psi=-\alpha_{1}(r,\phi_{1},\phi_{2},a_{1},a_{2}).
\end{equation}

\noindent We impose analogous boundary conditions on the surface of the second droplet.

The equations for~$\psi$ are solved exactly in Appendix~\ref{ADI}; in the far-field limit, the~$\psi$~contribution to the interaction energy~[i.e. the contribution from the first term of~Eq.~(\ref{NewEnergy})] reduces to

\begin{equation}
\label{FarFieldPairPotentialPsi}
V_{\psi}=\frac{\pi g(\alpha_{2}-\alpha_{1})^{2}}{\ln\left(\frac{r^{2}}{a_{1}a_{2}}\right)},
\end{equation}

\noindent where~$\alpha_{1}$~and~$\alpha_{2}$ are functions of~$r,a_{1},a_{2}$ and positions of the defects, which are specified by $\phi_{1}$~and~$\phi_{2}$ in the far-field limit. Equation~(\ref{FarFieldPairPotentialPsi}) has strong~$1/\ln(r)$ dependence, but as we will see later one can always arrange the defects in a way to set~$\alpha_{1}=\alpha_{2}$ and eliminate~$V_{\psi}$.

For the~$\chi$~field; Eqs.~(\ref{EquationPsiChi2})~and~(\ref{BoundaryConditionsChi}) could be solved approximately by the method of images. The number of images is infinite, but only a few of them contribute to the leading order. Apart from the~$p$ images inside each droplet that we considered in the previous section, we have to include images inside the second droplet of the whole structure in and around the first droplet~(defects, their images, and~$+2$~charge at the center of the first droplet; see Appendix~\ref{ASD}), as well as images inside the first droplet of the whole structure near the second droplet.\protect\footnote{\label{FMoreImages}Unlike Ref.~\cite{Pettey2}, we find that one has to consider more than~$2p$ images}  We use this solution to compute the energy, which we then Taylor expand to the leading order in powers~of~$1/R$. The result for the $\chi$~part of the droplet-droplet pair potential is

\begin{equation}
\label{FarFieldPairPotentialChi}
\begin{split}
V_{\chi}(r,\phi_{1},\phi_{2})_{\chi}=&(-1)^{p}\pi A(p)g\left(\delta^{p}+\frac{1}{\delta^{p}}\right)^{2}\\&\times\cos[p(\phi_{1}+\phi_{2})]\left(\frac{a_{1}^{p}a_{2}^{p}}{r^{2p}}\right),
\end{split}
\end{equation}

\noindent where~$A(p)$ is given in Table~\ref{Table}.

\begin{table}
\begin{ruledtabular}
\begin{tabular}{llll}
%\hline\hline
$p$ \hspace{3em} &  $A(p)$ \hspace{3em} & $r_{\rm{min}}/a$ \hspace{5em} & $\Delta/g$ \\
\hline
$1$ & $2$ & $2.878$ & $\infty$\\

$2$ & $3$ & $2.190$ & $1.229$ \\

$4$ & $35/2$ & $2.167$ & $0.996$\\

$6$ & $154$ & $2.11$ & $0.77$ \\
%\hline\hline
\end{tabular}
\end{ruledtabular}
\caption{Values of~$A(p)$ for~Eq.~(\protect\ref{FarFieldPairPotentialChi}), preferred separations~$r_{\rm{min}}$ and barriers to coalescence~$\Delta$ for physically relevant values~of~$p$~and~$a_{1}=a_{2}$. Values of~$r_{\rm{min}}/a$~and~$\Delta/g$ are accurate up to the last digit.}
\label{Table}
\end{table}

There are two important features of these far-field pair potentials. First, for any value of~$p$ there exist angles~$\phi_{1}$~and~$\phi_{2}$ such that~$V_{\chi}(r,\phi_{1},\phi_{2})<0$, and therefore~$V_{\chi}(r)$ can lead to an attraction, unlike~$V_{\psi}$, which is always repulsive. Second, there is a one-dimensional family of degenerate energies, because one can always add an arbitrary angle to~$\phi_{1}$ and subtract it from~$\phi_{2}$ without changing~$V_{\chi}$. Similarly,~$V_{\psi}$ depends only on the difference between~$\alpha_{1}$~and~$\alpha_{2}$.

To complete the analysis we only have to find~$\alpha_{1}$~and~$\alpha_{2}$ in terms of~$\phi_{1},\phi_{2}$,~and~$r$. In principle this dependence is given by the set of images described above but the explicit expressions are somewhat complicated. It is, however, easy to see that when~$a_{1}=a_{2}$~and~$\phi_{1}=\phi_{2} \; \rm{mod} \; 2\pi/p$ the contribution to the energy from the $\psi$~field should be small due to symmetry. Moreover, one can show analytically that for this orientation of the $p$-gons, $V_{\psi}$~falls of at least as~$r^{-(2p+1)}/\ln(r)$, which is much faster than~$V_{\chi}$. Thus, for~$a_{1}=a_{2}$ we can minimize the total interaction energy~$V=V_{\psi}+V_{\chi}$ by requiring that

\begin{equation}
\label{MinimumCondition1}
\phi_{1}=\phi_{2} \; \rm{mod} \; \left(\frac{2\pi}{p}\right),
\end{equation}

\begin{equation}
\label{MinimumCondition2}
\phi_{1}+\phi_{2}=\frac{[1+(-1)^{p}]\pi}{2p} \; \rm{mod} \; \left( \frac{2\pi}{p}\right),
\end{equation}

\noindent where Eq.~(\ref{MinimumCondition1}) follows from Eq.~(\ref{FarFieldPairPotentialChi}). If~$p=1$ the solution is~$\phi_{1}=\phi_{2}=0$ or $\phi_{1}=\phi_{2}=\pi$ and if~$p$ is even~$\phi_{1}=\phi_{2}=\pi/(2p)+\pi k/p$, where~$k$ is an integer. For different size droplets~Eq.~(\ref{MinimumCondition2}) still holds but~Eq.~(\ref{MinimumCondition1}) must be generalized to an implicit equation

\begin{equation}
\label{MinimumCondition3}
\alpha_{1}(r,a_{1},a_{2},\phi_{1},\phi_{2})=\alpha_{2}(r,a_{1},a_{2},\phi_{1},\phi_{2}).
\end{equation}

\noindent Generically, Eqs.~(\ref{MinimumCondition2})~and~(\ref{MinimumCondition3}) have a solution, which minimizes both~$V_{\psi}$~and~$V_{\chi}$. The large~$r$ pair potential for this favorable relative orientation is then attractive and is given by

\begin{equation}
\label{FarFieldPairPotential}
V(r)=-\pi A(p)g\left(\delta^{p}+\frac{1}{\delta^{p}}\right)^{2}\left(\frac{a_{1}^{p}a_{2}^{p}}{r^{2p}}\right).
\end{equation}

It is worth noting that~$V_{\chi}$ obtained here agrees with Ref.~\cite{Pettey2} for~$p=1$ considered therein, but we have a different expression for~$V_{\psi}$. We think that our result is more physical because~$V_{\psi}$ is repulsive, does not depend on the system size, and decays for large~$r$, unlike the potential calculated in Ref.~\cite{Pettey2}. Our result also generalizes the work of Tasinkevych, Silvestre, Patr\'{i}cio, and Telo~da~Gama~\cite{Tasinkevych}, who considered the case of~$p=2$,~$a_{1}=a_{2}$, and~$\phi_{1}=\phi_{2}$. Since in this case~$V_{\psi}=0$, the authors did not discuss the role of~$V_{\psi}$ in the pairwise interaction. The superposition principle does \textit{not} hold for reasons related to footnote~\ref{FMoreImages}; as a result, the pair potential in Ref.~\cite{Tasinkevych} differs from ours in its dependence on~$\delta$.

\section{\label{SNCPI}Numerical calculation of pairwise interaction}

\begin{figure}
\includegraphics[width=\columnwidth]{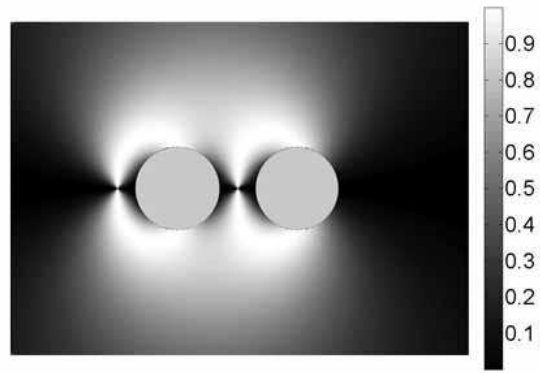}
\caption{Equilibrium texture for a 2D,~$p=1$,~liquid crystal for equally sized droplets at the preferred separation, with free boundary conditions at infinity. A defect with topological charge~$-1$~appears just to the left of each droplet. The brightness represents the square of sine of the order parameter, which should be similar to what one would observe with crossed polarizers.}
\label{FigureTexture}
\end{figure}

As the droplets get closer to each other and subleading terms become important, we expect the interaction between the clouds of defects to give rise to a repulsive force somewhat similar to electron mediated repulsion between atoms of ordinary matter. To check this conjecture we must extend our calculation~to~$r\approx a_{1},a_{2}$. To that end we devised an algorithm of solving Eq.~(\ref{EquationPsiChi2}) subject to the boundary conditions~(\ref{BoundaryConditionsChi}) and specified positions of the defects; we used Eq.~(\ref{PsiEnergy}) from Appendix~\ref{ADI} to account for the energy of the $\psi$~field. This procedure was carried out both by solving Laplace's equation on a lattice and by iterative evaluation of a large number of images. The results of these two methods were mutually consistent and the relative error of energy could be made as low~as~$10^{-5}$. An output of one such calculation for~$p=1$ is shown in Fig.~\ref{FigureTexture}. We then used a downhill simplex method~\cite{Nelder} to minimize the free energy with respect to the positions of the topological defects; different starting simplexes were used to check that the algorithm converges to the correct global minimum. The procedure was repeated for several separations~$r$ until a reliable graph of the pair potential was obtained. This calculation was done for~$p=1,2,4,6$ and different ratios~of~$a_{2}/a_{1}$.

\begin{figure}
\includegraphics[width=\columnwidth]{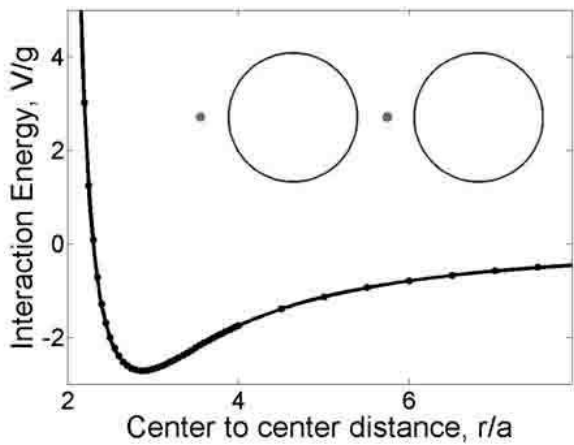}
\caption{Pair potential for~$p=1$~and~$a_{1}=a_{2}$. The dots on the curve are the actual data points while the line is just a guide to the eye. The inset shows equilibrium locations of the two defects at the preferred separation, i.e. at the minimum of the pair potential. The defects prefer to lie on the center to center line, as shown on the inset, for all values of~$r$. The behavior for other values of~$a_{2}/a_{1}$ is qualitatively similar.}
\label{PairPotentialp1}
\end{figure}

A result of one such calculation for~$p=1$,~$a_{2}/a_{1}=1$ is shown in Fig.~\ref{PairPotentialp1}, where the inset shows locations of the defects at the minimum of the pair potential. At large separations, droplets attract~and~$V(r)=-9\pi g(a/r)^{2}$ in agreement with~Eq.~(\ref{FarFieldPairPotential}). Note, there is no logarithmic contribution from~$V_{\psi}$ because~$\alpha_{1}=\alpha_{2}$ for the equilibrium configuration of defects. At small separations,~$V(r)$ is repulsive and when the droplets touch it tends to infinity, thus inhibiting coalescence. A distinct feature of this pair potential is a minimum which implies that at sufficiently low temperatures two droplets come together to form a stable dimer, as shown in Fig.~\ref{FigureTexture}.

As we showed in the previous section, the defects for~$p=1$ prefer to lie on the line connecting the centers of the circles when the droplets are far apart (see the inset in Fig.~\ref{PairPotentialp1}.) From our simulation we find that this behavior persists for all separations,~$r$ and relative droplet sizes studied~$a_{2}/a_{1}$~(we explored~$0.1\leq a_{2}/a_{1} \leq 1$), because for this configuration of the defects~$\alpha_{1}=\alpha_{2}$~and~$V_{\psi}$ is zero. Therefore, even when the droplets come very close there is a defect sitting between them, which naturally leads to a diverging pair potential as~$r\rightarrow(a_{1}+a_{2})$. This divergence will be cut off either by the core radius, or when our strong anchoring boundary conditions are relaxed. Textures and~$r_{\rm{min}}$ obtained here agree reasonably well with measurements by Cluzeau \textit{et al.}~\cite{Cluzeau1,Cluzeau2} for~a~$p=1$ system and with simulations in Refs.~\cite{Tasinkevych,Patricio} for~$p=1$~and~$p=2$ systems.

\begin{figure}
\includegraphics[width=\columnwidth]{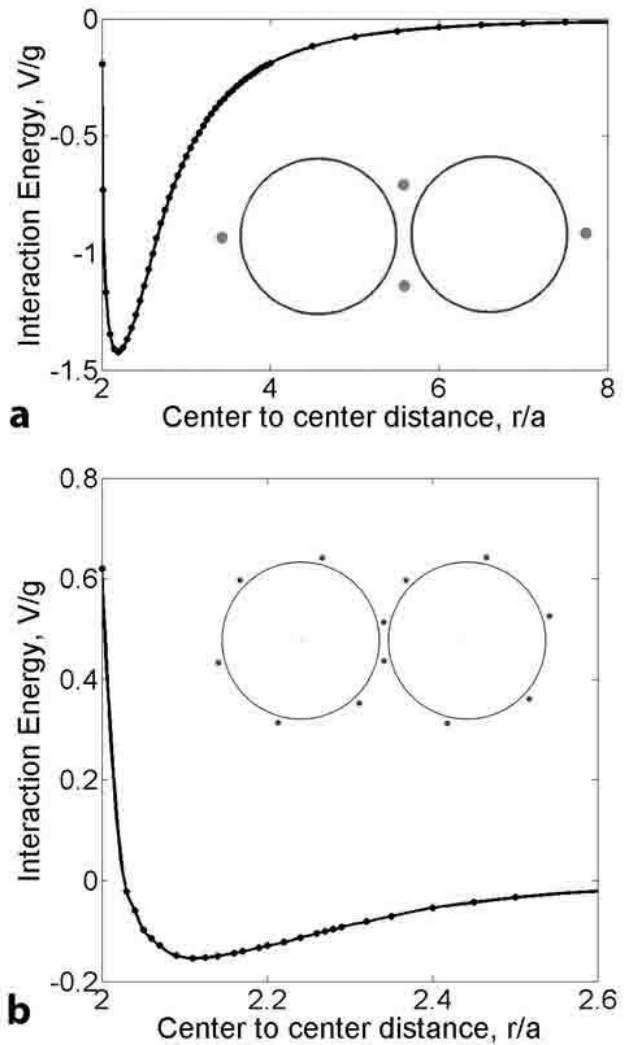}
\caption{Pair potentials for~(a)~$p=2,a_{1}=a_{2}$~and~(b)~$p=6$,~$a_{1}=a_{2}$. The dots on the curves are the actual data points while the lines are just guides to the eye. The insets show equilibrium locations of the defects at the preferred separation, i.e. at the minimum of the pair potential. The behavior for other values of~$a_{2}/a_{1}$ is qualitatively similar. Note the centrosymmetric configuration of the defects.}
\label{PairPotentialp2}
\end{figure}

For~$p=2,4,$~and~$6$ the pair potential also has a minimum, but~$V(r)$ remains finite when droplets touch. The difference arises because there is a centrosymmetric configuration of the defects that minimizes~$V_{\chi}$ and eliminates~$V_{\psi}$. In addition, this configuration~(see the insets of~Fig.~\ref{PairPotentialp2}) does not have a defect between the droplets, the cause of the divergence for~$p=1$. Examples of such pair potentials are shown in Fig.~\ref{PairPotentialp2} for~$p=2,6$,~and~$a_{2}/a_{1}=1$.

\begin{figure}
\includegraphics[width=\columnwidth]{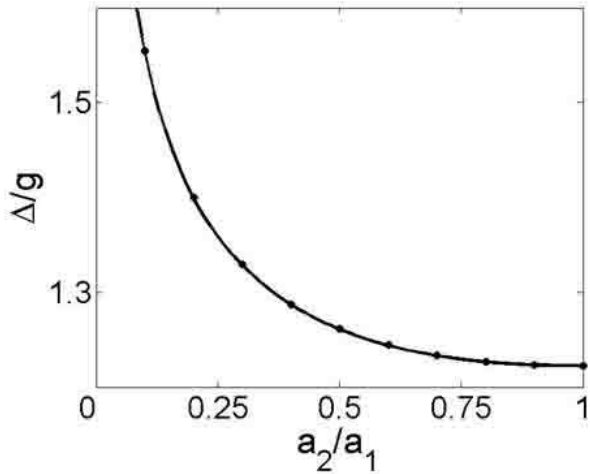}
\caption{Barriers to coalescence for~$p=2$ as a function of the ratio of droplet sizes,~$\Delta(a_{2}/a_{1})$. The dots on the curve are the actual data points while the line is just a guide to the eye.  Note that~$\Delta(a_{2}/a_{1})=\Delta(a_{1}/a_{2})$.}
\label{FigureDelta}
\end{figure}

We quantify barriers to coalescence by introducing a free energy difference~$\Delta$ between states when droplets touch~($r=a_{1}+a_{2}$) and when they are at the preferred separation~$r_{\rm{min}}$, which provides a lower bound on the activation barrier, and leads to~$e^{-\Delta/k_{B}T}$-fold decrease in the rate of coalescence. It is important to know how~$\Delta$ depends on the ratio~$a_{2}/a_{1}$ of the two droplet sizes to understand the dynamics of coalescence. This dependence is summarized in Fig.~\ref{FigureDelta} for~$p=2$; we find similar behavior for~$p=2,4,$~and~$6$. One can see that the more asymmetric the size ratio the higher the barriers to coalescence, which implies almost equally sized droplets are more likely to coalesce than droplets of different sizes.

For all physically relevant values~of~$p$ and ratios~of~$a_{2}/a_{1}$ studied, we found barriers to coalescence and a minimum of the pair potential, which implies topological stabilization and dimerization of isolated droplet pairs. Table~\ref{Table}~(see Sec.~\ref{SFFI}) summarizes preferred separations and minimum barriers to coalescence for different values~of~$p$~and~$a_{1}=a_{2}$.

\begin{figure}
\includegraphics[width=\columnwidth]{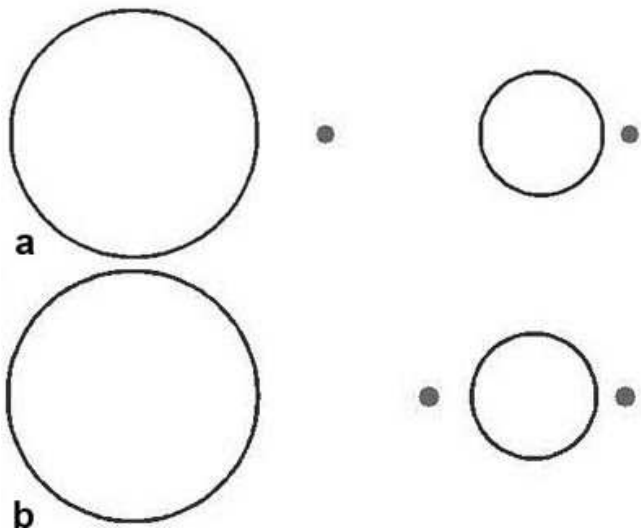}
\caption{The smaller droplet annexes a defect as the separation between the circles decreases. Both (a) and (b) show equilibrium configurations of the defects for two droplets with~$a_{1}=2a_{2}$. In (a) $r=3.3a_{1}$, and each droplet has a companion defect, while in (b) $r=3.2a_{1}$, and both defects are located near the smaller droplet. The figure shows annexing for~$p=1$, but similar behavior has been observed for~$p=2$,~$4$~and~$6$.}
\label{FigureAnnexing}
\end{figure}

The formation of droplet dimers has a number of common features with the chemistry of diatomic molecules. For equal droplet sizes the defects are divided evenly between the droplets similarly to a nonpolar covalent bond. If the sizes are slightly different, the defects shift toward the smaller droplet as in a polar covalent bond. Eventually, when the radii differ by about a factor of 2 or more, the smallest droplet annexes one of the defects and the bond becomes ionic (see Fig.\ref{FigureAnnexing}).

Our calculations not only explain the stability of emulsions, but also suggest an experimental signature of topological emulsions. For example, hexatic order can be difficult to detect directly, because it does not couple directly to the polarization of light; detection via x-ray scattering experiments on two-dimensional films can be challenging. In contrast, dimerization is observable by ordinary light microscopy and our theory predicts a \textit{universal} equilibrium separation that depends only on the sizes of the droplets but not on the properties of the material. Formation of dimers with separation that agrees with the theory presented here~(there are no adjustable parameters) would be a strong indicator of the presence of $p$-fold order surrounding coexisting droplets. In addition, our calculation of the pair potential could be used to extract the stiffness constant~$g$ from experimental measurements of thermal fluctuations about the preferred separation. This signature could also be important for the order parameters~($p=1$ and $p=2$) that couple directly to light, because precise measurements of the texture and location of defects could be more difficult than measurements of the separation between the droplets~\cite{Cluzeau2}.

\section{\label{SChaining}Chaining}

Even though understanding the interaction of two droplets is important, real systems can be more subtle, because simple pairwise interactions sometimes lead to complex behavior of matter as a whole. Indeed, many experiments report self-assembly of droplets into chains at low concentrations and lattices at high concentrations~\cite{Voltz,Cluzeau1,Cluzeau2,Dolganov}. Chaining could be explained by  texture-induced dipole-dipole interactions in the presence of an in-plane aligning field due to boundary conditions, similar to ferrofluids~\cite{Larson}. However, this explanation seems insufficient in two ways. First, at small separations, as one might expect in chains, the interaction is no longer dipole-dipole as we showed above. Second, the explanation appeals to an aligning field or aligning boundary conditions at infinity, dismissing a possibility of chaining in their absence.

Unfortunately, our numerical method of images becomes prohibitively complex for more than a few droplets and requires free boundary conditions. In addition, for several droplets one has to solve for~$\psi$ numerically. Nevertheless, there are two important cases where the progress can be made.

The first case is chaining for~$p=1$, where we know from experiments by Cluzeau \textit{et al.}~\cite{Cluzeau1} that defects lie along the chain between the droplets. For this configuration~$V_{\psi}$ is zero (which probably drives the chaining) and we can find distance between neighbors in a chain of equal droplets, which is~$2.89a$.

\begin{figure}
\includegraphics[width=\columnwidth]{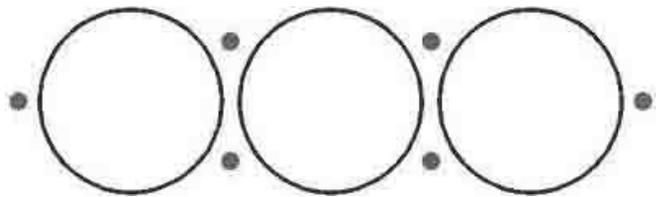}
\caption{Minimal energy configuration of three equally sized droplets,~$p=2$. The inter droplet spacing is~$2.19a$, which is very close to the size of a dimer.}
\label{p2Chain}
\end{figure}

The second case is chaining for~$p=2$, where the texture around three and four equal droplets that minimizes~$V_{\chi}$ happens to set~$V_{\psi}$ to zero as well. This configuration of defects is shown in Fig.~\ref{p2Chain} and is analogous to a polymer such as polyethylene in chemistry. In fact, one can view the sealing defects on the ends as free radicals that facilitate the reaction of chain polymerization. An important conclusion is that in this case chaining is driven solely by the interaction of topological defects -- no external field is required.

\section{\label{STF}Effects of thermal fluctuations and boundary conditions}

It is well known that thermal fluctuations are especially important in two dimensions. In this section we discuss how they modify the results presented so far. In the one elastic constant approximation, considered in this paper, the effects of temperature depend on only one dimensionless parameter,~$\varkappa=k_{B}T/g$. Even though the temperature can be varied only in a limited range between neighboring phase transitions, in most liquid crystals,~$g$~can be changed by about two orders of magnitude by varying film thickness in the experiments of Ref.~\cite{Cluzeau1}. In fact,~$g \approx k_{B}T_{c}N$, where~$T_{c}$ is the temperature of the ordering phase transition in a single layer, and~$N$ is the number of layers in the film~\cite{Pettey2}. Thus one can roughly interpret~$\varkappa^{-1}$ as~$N$.

Even though thermal fluctuations lead to many important effects in liquids with a continuous \textit{XY} symmetry, below the Kosterlitz-Thouless transition most of them can be absorbed into a renormalized stiffness~$g$ that now depends not only on temperature but also on the length scale~\cite{Kosterlitz}. We neglect this dependence because it is rather weak, but we allow the defect positions to fluctuate in space according to the Boltzmann distribution. Then we compute thermal averages of interest either by direct numerical integration  weighted by Boltzmann factors, or by Monte Carlo simulations.

\begin{figure}
\includegraphics[width=\columnwidth]{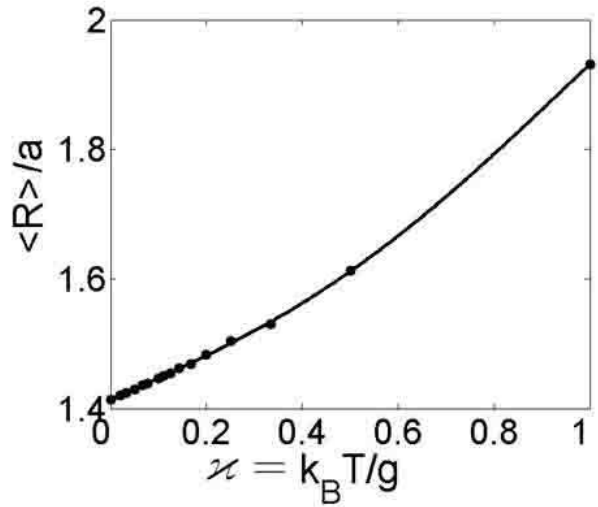}
\caption{Average separation between the droplet and the defect as a function of temperature for~$p=1$. The separation approaches a value given by~Eq.~(\protect\ref{Delta}) as~$T\rightarrow 0$.}
\label{TemperatureR}
\end{figure}

First we analyze a single droplet for~$p=1$. Due to the highly asymmetric shape of the potential in Fig.~\ref{PairPotentialp1} we expect that on average the defect is located further from the droplet than~Eq.~(\ref{Delta}) predicts. This effect is significant even for small values of~$\varkappa$ as shown in Fig.~\ref{TemperatureR}, and could be observable experimentally.

\begin{figure}
\includegraphics[width=81mm]{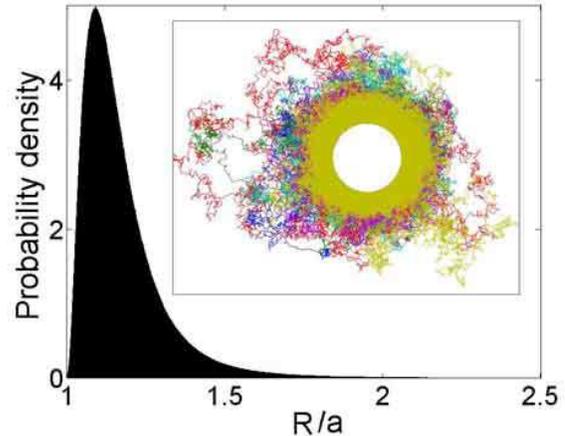}
\caption{ (Color online) Probability density  for the separation between a defect and the droplet for~$p=6$. The inset shows that the six defects execute random walks; different colors represent different defects. The yellowish annulus around the boundary arises because random walks overlap near the droplet, where the density of the defects is very high. The simulation was done at~$\varkappa^{-1}=25$ with~$10^6$ Monte Carlo steps, but only each tenth step is shown for clarity.}
\label{TemperatureOrbit}
\end{figure}

Effects of thermal fluctuations are even more dramatic for higher values of~$p$. Figure~\ref{TemperatureOrbit} shows the probability density for the separation between a defect and the droplet when~$p=6$; the inset shows random walk of the six defects. Similar to the blurred electron wave function of atomic physics, topological defects at high temperatures form a ``defect cloud'' around the droplet rather than a simple hexagon.

\begin{figure}
\includegraphics[width=\columnwidth]{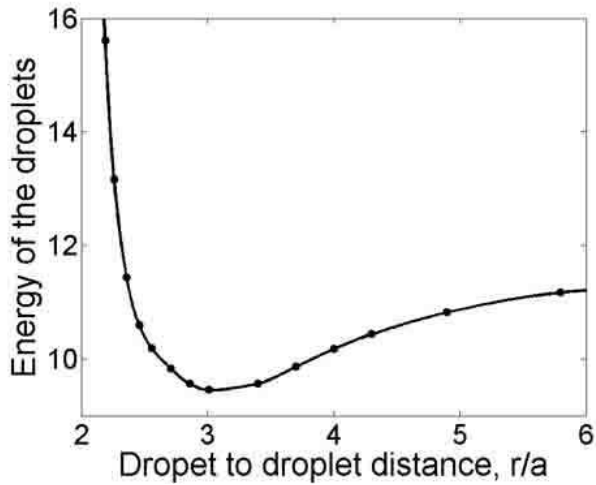}
\caption{Thermally renormalized pair potential for~$p=1$,~$a_{1}=a_{2}$,~and~$\varkappa^{-1}=4$. The dots on the curve are the actual data points while the line is just a guide to the eye. Energy of the droplets is measured in units of~$g$. Unlike Figs.~\protect\ref{PairPotentialp1}~and~\protect\ref{PairPotentialp2}, we did not subtract the energy of two isolated droplets.}
\label{TemperaturePairPotential}
\end{figure}

For two droplets, the quantity of interest is an effective pair potential for nonzero temperatures. By effective potential we mean thermal average of the bare pair potential with respect to the positions of the defects. Figure~\ref{TemperaturePairPotential} shows this quantity for~$p=1$ and~$\varkappa^{-1}=4$. Upon recalling that~$\varkappa^{-1} \approx N$, one can see that even for a film of only four layers thick there is still a minimum and barriers to coalescence, which suggests that our results are robust to thermal fluctuations. It is also interesting that~$r_{\rm{min}}$ increases with temperature~(and~decreases with film thickness). The trend is the opposite of what was found experimentally in Ref.~\cite{Cluzeau2}, suggesting that some other effect must account for the observations. Similar results were obtained for~$p=2$.

\section{Conclusions}
We have shown that two-dimensional dispersions of isotropic phase in a liquid with a continuous symmetry \textit{XY}-like order parameter are topologically stable, provided there are strong anchoring boundary conditions on the droplets. We have calculated the texture for one and two droplets and determined the pair potential for~$p=1,2,4,$~and~$6$. A cloud of defects surrounding each droplet insures topological charge neutrality and contributes an additional term to the free energy, which depends logarithmically on the size of the droplet.  At large separations, droplets of isotropic phase attract while at small ones they repel. The defect-mediated pair potential clearly indicates kinetic barriers to coalescence and favors dimerization. We developed a fast numerical method for obtaining the pair potential for droplets of similar and disparate sizes and used it to calculate activation barriers and the sizes of dimers. The latter results could be used to test experimentally for the presence of ``hidden'' order parameters with, e.g.,~$p=4$~and~$p=6$. We also analyzed chaining and showed that it could be driven not only by an aligning field due to the boundary conditions~\cite{Poulin,Lubensky} but instead solely by the interaction between the droplets and topological defects for~$p=2$. Finally, we addressed the effects of thermal fluctuations on the texture around a single droplet and on the pair potential and indicated how the preferred separation could be affected by the thickness of the film.

\begin{acknowledgments}
 One of us~(D.R.N.) acknowledges extensive conversations with F.~J\"{u}licher and M.~Gopalakrishnan while enjoying the hospitality of the Max Plank Institute for Complex Systems in Dresden, Germany, during the early part of this investigation. He would also like to thank T.~Baumgart and W.W.~Webb for introducing him to the problem of rafts in lipid membranes. We also thank D.~Nurgaliev for useful discussions. This work was supported by the National Science Foundation, through Grant No. DMR-0654191 and the Harvard Materials Research Science and Engineering Center through Grant No. DMR-0213805.
\end{acknowledgments}

\appendix

\section{\label{ASD}Energy of a single droplet}

In this appendix we derive Eqs.~(\ref{SingleDropletEnergy})~and~(\ref{Delta}) for the energy and position of the satellite defects around a droplet, starting from just before Eq.~(\ref{SingleDropletSolution}). To display the solution, we first introduce some notation. Let~$A_{i},~i=1,\ldots,p$ be the positions of the defects created in the liquid crystal, and~$A'_{i},~i=1,\ldots,p$ their images in the circle~$\mathcal{C}$ of radius~$a$ centered~at the origin~$O$, which represents the surface of the droplet. The images are obtained by inversion, i.e. the image~$(x',y')$ of~$(x,y)$ in~$C$ is given by

\begin{equation}
\label{Inversion}
\begin{aligned}
x'&=\frac{xa^{2}}{x^{2}+y^{2}},\\
y'&=\frac{ya^{2}}{x^{2}+y^{2}}.\\
\end{aligned}
\end{equation}

\noindent At each position~$A_{i}$~we define an angular variable~$\phi_{i}(X)$, which is the angle~between axis~$OA_{i}$ and vector~$\overrightarrow{A_{i}X}$, where~$X$ is any point in the two-dimensional space. Similarly, we define angular variables~$\phi'_{i}$~at~$A'_{i}$. We also introduce~$\phi_{0}(X)$~as the angle between axis~$OA_{1}$~and~vector~$\overrightarrow{OX}$. The solution in terms of these new variables is given by

\begin{equation}
\theta=2\phi_{0}-\frac{1}{p}\sum_{i=1}^{p}(\phi_{i}+\phi'_{i})+\frac{\pi}{p}.
\label{EqSolution}
\end{equation}

\noindent Equation~(\ref{EqSolution}) satisfies~Eq.~(\ref{LaplaceEquation}) because~$\theta(x,y)=\arctan \left(\frac{y-y_{0}}{x-x_{0}}\right)$ is a solution of Laplace's equation for arbitrary~$(x_{0},y_{0})$. It also satisfies the strong anchoring boundary condition on the droplet due to a simple geometrical fact~(see Fig.~\ref{CircleDrawing}):

\begin{figure}
\includegraphics[width=\columnwidth]{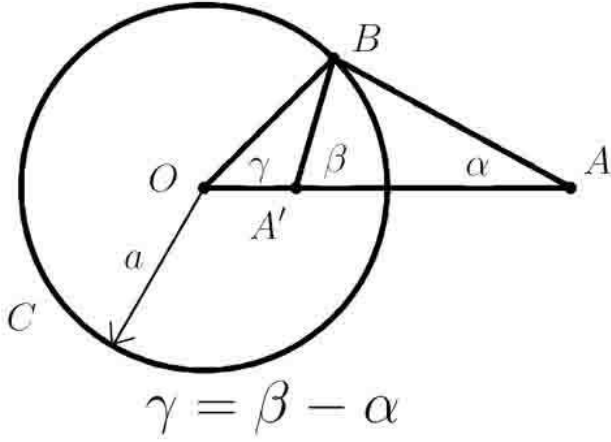}
\caption{Illustration of the geometrical fact that~$\angle BOA =\angle BA'A - \angle OAB$. }
\label{CircleDrawing}
\end{figure}

\begin{quote}

\itshape

If~$\mathcal{C}$ is a circle of radius~$a$ centered at point~$O$,~$A$~is~any point outside~$\mathcal{C}$~and~$A'$~is the image~of~$A$ under inversion, and $B$ is any point~on~$\mathcal{C}$, then~$\angle BOA =\angle BA'A - \angle OAB$.

\end{quote}

\noindent With this observation, we conclude that at the boundary,~$\phi_{k}+\phi'_{k}=\phi_{0}+\pi-2\pi(k-1)/p$. Then~$\theta$~at~the~boundary is given by~$\theta=\phi_{0}+\pi/p-1/p[p\pi-\pi(p-1)]=\phi_{0}$, i.e., the boundary conditions are satisfied.

To calculate the free energy it is advantageous to introduce another field~$\chi$, such that~$\partial_{i}\chi=\bm{\epsilon}_{ik} \partial_{k} \theta$,~[see Eq.~(\ref{PsiChiDecomposition})], where~$\bm{\epsilon}_{ik}$~is the antisymmetric tensor in two dimensions,~$\bm{\epsilon}_{xx}=\bm{\epsilon}_{yy}=0$, $\bm{\epsilon}_{xy}=-\bm{\epsilon}_{yx}=1$. It is easy to see that the free energy in terms of the Cauchy conjugate function~$\chi$ has the same functional form as~Eq.~(\ref{GradientExpansionEnergy}),

\begin{equation}
 F=\frac{g}{2}\int(\protect\bm{\nabla}\chi)^{2}d^{2}x.
 \label{EqEnergyChi}
\end{equation}

\noindent The solution in terms of~$\chi$ is given by

\begin{equation}
\begin{split}
\chi(\mathbf{r})=&2 \ln|\mathbf{r}|-\frac{1}{p}\sum_{i=1}^{p}(\ln|\mathbf{r}-\overrightarrow{OA_{i}}|+ \ln|\mathbf{r}-\overrightarrow{OA'_{i}}|)
\end{split}
\label{EqSolutionChi}
\end{equation}

\noindent Upon integrating by parts in~Eq.~(\ref{EqEnergyChi}), we find

\begin{equation}
F=-\frac{g}{2}\int_{\mathcal{C}}\chi (\mathbf{\protect\bm{\nabla}} \cdot \chi \mathbf{\hat{n}})dl - \frac{g}{2}\int \chi \protect\bm{\nabla}^{2} \chi d^{2}x
\label{EqFreeEnergyTwoTerms}
\end{equation}

\noindent where~$\mathbf{\hat{n}}$ is an outward unit normal on the circle and the contour~$C$ includes both the circle and a boundary at infinity. It is clear that for our solution the surface integral at infinity vanishes.

Note that in polar coordinates around the center of the circle~$(r,\phi_{0})$, ~$\protect\bm{\nabla}\theta(\mathbf{r})=\mathbf{\hat{n}}\frac{\partial\theta}{\partial r}+\frac{\bm{\hat{\phi_{0}}}}{r}\frac{\partial\theta}{\partial \phi_{0}}$. Therefore~$\protect\bm{\nabla}\chi(\mathbf{r})=\frac{\mathbf{\hat{n}}}{r}\frac{\partial\theta}{\partial \phi_{0}}-\bm{\hat{\phi_{0}}}\frac{\partial\theta}{\partial r}$, because by definition,~$\protect\bm{\nabla} \chi$~is~$\protect\bm{\nabla}\theta$ rotated~by~$-\pi/2$. Since on the boundary~$\theta=\phi_{0}$, we have

\begin{equation}
\int_{\mathcal{C}}\chi (\mathbf{\protect\bm{\nabla}} \cdot \chi \mathbf{\hat{n}}) dl=\int_{\mathcal{C}}\chi d\phi_{0}.
\end{equation}

\noindent This equation can be further simplified with the help of the identity

\begin{equation}
\label{EqIntLn}
\int_{0}^{2\pi}\frac{\ln|u^2+v^2-2 u v \cos\phi|}{\ln(\max\{|u|,|v|\})} d\phi=4\pi ,
\end{equation}

\noindent which leads to

\begin{equation}
\int_{\mathcal{C}}\chi (\protect\bm{\nabla} \chi \mathbf{\hat{n}}) dl=-2\pi \ln\left(\frac{R}{a}\right).
\label{OneDropTerm1}
\end{equation}

To evaluate the second term~in~eq.~(\ref{EqFreeEnergyTwoTerms}) we use~$\protect\bm{\nabla}^{2}   \ln|\mathbf{r}-\mathbf{r_{0}}|=2\pi\delta^{(2)}(\mathbf{r}-\mathbf{r_{0}})$ and the fact that all defects contribute equally due to symmetry. Thus up to core energies that depend on microscopic details we find

\begin{equation}
\begin{split}
\int \chi \protect\bm{\nabla}^{2} \chi d^{2}x=& -4\pi \ln(R) + \frac{2\pi}{p} \ln\left(R-\frac{a^{2}}{R}\right) + \\& \frac{2\pi}{p} \sum_{i=2}^{p}\left(\ln|\overrightarrow{OA_{i}}-\overrightarrow{OA_{1}}|+\right.\\&\qquad\quad\;\;\left.\ln|\overrightarrow{OA'_{i}}-\overrightarrow{OA_{1}}|\right).
\end{split}
\end{equation}

We simplify this formula by taking advantage of complex numbers: let~$\overrightarrow{OA_{i}}=R z_{i}$ and~$\overrightarrow{OA'_{i}}=\frac{a^{2}}{R} z_{i}$, where~$z_{k}=\exp[ 2 \pi i (k-1)/p]$. Then we use the following chain of identities; valid for~$z_{1}\equiv 1$:
\begin{equation}
\begin{split}
 \prod_{i=1}^{p}(z-z_{i})&=z^{p}-1 \Rightarrow, \\
 \prod_{i=2}^{p}(z-z_{i})&=\sum_{i=0}^{p-1}z^{i} \Rightarrow, \\
 \prod_{i=2}^{p}(z_{1}-z_{i})&=\sum_{i=0}^{p-1} 1=p,
\end{split}
\end{equation}

\noindent to sum up the terms in the previous equation, which reduces to

\begin{equation}
\label{DefectsTerm}
-\int \chi \protect\bm{\nabla}^{2} \chi d^{2}x= \frac{\pi}{p}\ln\left(\frac{R^{2p+1}}{p(R^{2p}-a^{2p})c}\right)+pE_{c},
\end{equation}

\noindent where we put the core energy~$E_{c}$ back in and introduced the core radius~$c$. Upon combining the Eqs.~(\ref{DefectsTerm})~and~(\ref{OneDropTerm1}) we obtain the free energy~(\ref{SingleDropletEnergy}), which upon minimization gives~Eq.~(\ref{Delta}).

\section{\label{ASDI}Ordered droplets in disordered fluid}

In this appendix we briefly discuss the equilibrium configuration of defects \textit{inside} a droplet of anisotropic two-dimensional liquid embedded in disordered host fluid. As before, we assume strong anchoring boundary conditions, which imply that the net charge of the defects in the interior must sum up to~$+1$. Therefore we need~$p$ point defects with charges~$1/p$. It is natural to assume that these can be arranged around the center of the droplet either in regular $p$-gon or in regular $(p-1)$-gon with a defect in the center. We can then calculate preferred separation between the defects and the center of the circle,~$R=\delta a$. This can be done as in~Appendix~\ref{ASD}, and the results are given below.

\noindent Regular $p$-gon,

\begin{equation}
F=-\frac{\pi g}{p} \ln\left(\frac{(a^{2p}-R^{2p})R^{p-1} p c}{a^{3p}}\right),
\end{equation}

\noindent which upon minimization gives

\begin{equation}
\delta=\left(\frac{p-1}{3p-1}\right)^{\frac{1}{2p}}.
\end{equation}

\noindent Regular $(p-1)$-gon~(with central defect),

\begin{equation}
\begin{split}
F=&-\frac{\pi g(p-1)}{p^{2}} \ln\left(\frac{(a^{2p-2}-R^{2p-2})R^{p} (p-1)}{a^{3p-2}}\right)+\\&\frac{\pi g}{p}\ln\left(\frac{a}{c}\right),
\end{split}
\end{equation}

\noindent which leads to

\begin{equation}
\delta=\left(\frac{p}{3p-2}\right)^{\frac{1}{2p-2}}.
\end{equation}

Our calculation shows that the $p$-gon is energetically favorable compared to a~$(p-1)$-gon for all values of~$p$. Depending on anchoring energy, it is also possible to have surface and virtual defects instead of topological ones, but we do not pursue this possibility further here (see Ref.~\cite{Pettey} for an analysis when~$p=1$).

\section{\label{ADI}Solution for $\psi-$field}

In this appendix we calculate the droplet-droplet potential contribution~$V_{\psi}$. We assume that droplets of radii~$a_{1}$~and~$a_{2}$ are located at~$(0,0)$~and~$(r,0)$ respectively, and that~$\psi$ satisfies Laplace's equation everywhere in space. $\psi$ is forced to equal to~$-\alpha_{1}$~and~$-\alpha_{2}$ on the surface of the droplets. This problem can be solved by taking advantage of the fact that conformal mappings leave Laplace's equation invariant. Hence, we search for a conformal mapping that transforms two circles~(representing the droplet boundaries) into an annular ring at the origin, where Laplace's equation can be solved easily. Upon passing to complex coordinates~$\xi$~and~$z$, this transformation is

\begin{equation}
\label{ConformalMapping}
\xi=\frac{a_{1}^{2}-wz}{z-w},
\end{equation}

\noindent where the real parameter~$w$ is given by

\begin{equation}
\label{EquationW}
\begin{split}
w=& \frac{r^{2}+a_{1}^{2}-a_{2}^2}{2r} - \\  & \frac{\sqrt{(a_{1}^{2}-a_{2}^{2})^{2}+r^{2}[r^{2}-2(a_{1}^{2}+a_{2}^{2})]}}{2r}.
\end{split}
\end{equation}

\noindent The outer radius of the ring is~$a_{1}$ and the inner radius~$\tilde{a}$ is

\begin{equation}
\label{InnerRadius}
\tilde{a}=\frac{a_{1}^2-w(r-a_{2})}{r-a_{2}-w}.
\end{equation}

\noindent The solution of Laplace's equation in this annular domain is then given by

\begin{equation}
\label{PsiSolution}
\psi=\frac{\alpha_{2}\ln(|\xi|/a_{1})-\alpha_{1}\ln(|\xi|/\tilde{a})}{\ln(a_{1}/\tilde{a})},
\end{equation}

\noindent which can be easily expressed in terms of~$z=(x,y)$ by inverting Eq.~(\ref{ConformalMapping}). Now,~$V_{\psi}=\frac{g}{2}\int(\protect\bm{\nabla}{\psi})^{2}d^{2}x$ can be calculated as follows:

\begin{equation}
\label{PsiEnergy}
\begin{split}
V_{\psi}=&\frac{g\alpha_{1}}{2}\int_{droplet\;one}\protect\bm{\nabla}\psi\cdot\mathbf{\hat{n}}dl-\frac{g\alpha_{2}}{2}\int_{droplet\;two}\protect\bm{\nabla}\psi\cdot\mathbf{\hat{n}}dl\\=&\frac{g\alpha_{1}}{2}\int_{droplet\;one}\protect\bm{\nabla}^{2}\psi d^{2}x-\frac{g\alpha_{2}}{2}\int_{droplet\;two}\protect\bm{\nabla}^{2}\psi d^{2}x\\=&\frac{\pi g(\alpha_{1}-\alpha_{2})^{2}}{\ln(a_{1}/\tilde{a})}.
\end{split}
\end{equation}

\noindent In the limit~$r\gg a_{1},a_{2}$ this equation reduces to~Eq.~(\ref{FarFieldPairPotentialPsi}).

\bibliographystyle{apsrev}

\end{document}